\documentclass[conference,10pt]{IEEEtran}
\pdfoutput=1

\usepackage{graphicx}
\usepackage{epstopdf}
\usepackage{dblfloatfix}
\usepackage{booktabs}
\usepackage{enumitem}
\usepackage{multirow}
\usepackage{url}
\usepackage[normalem]{ulem}
\usepackage{color}
\usepackage{cite}
\usepackage{caption} 
\usepackage{amssymb}
\usepackage[cmex10]{amsmath}
\usepackage{gensymb}
\usepackage{amsthm}
\usepackage[numbers,square,comma]{natbib}

\begin{document}
	
\bstctlcite{IEEEexample:BSTcontrol}




\title{Power and Thermal Analysis of Commercial Mobile Platforms: Experiments 
and Case Studies}
	
\author{\IEEEauthorblockN{Ganapati Bhat$^1$, Suat Gumussoy$^2$, and Umit Y. 
Ogras$^1$}
	\IEEEauthorblockA{$^1$School of Electrical Computer and Energy Engineering, 
	Arizona State University, Tempe, AZ\\
	$^2$IEEE Member \\
		Email: gmbhat@asu.edu, suat@gumussoy.net, umit@asu.edu}
}


\maketitle
\begin{abstract}
State-of-the-art mobile processors can deliver fast response time and high 
throughput to maximize the user experience. 
However, high performance comes at the expense of larger power density, 
which leads to higher skin temperatures.
Since this can degrade the user experience, there is a strong need for power consumption and thermal analysis in mobile processors.  
In this paper, we first perform experiments on the Nexus 6P phone to study the power, 
performance and thermal behavior of modern smartphones. 
Using the insight from these experiments, we propose a control algorithm that 
throttles select applications without affecting other apps. 
We demonstrate our governor on the Exynos 5422 processor employed in the 
Odroid-XU3 board.
\end{abstract}

\section{Introduction}

Commercial mobile platforms need to support multiple 
application categories, 
such  as games, navigation, social media apps, and video players. 
%
In order to deliver competitive performance while running these apps, 
state-of-the-art mobile processors integrate multiple powerful general-purpose 
cores, graphics processing cores~(GPU), and accelerators~\cite{Qual_Snap810,gupta2016generic}. 
%
%
At the same time, mobile platforms are severely limited by thermal design power 
(TDP) due to their small size and lack of active 
cooling~\cite{pagani2014tsp,raghavan2012computational,
	shafique2014eda,gupta2017dynamic}.
Power dissipation increases not only the junction temperature on the chip
but also the skin temperature of the platforms, 
which directly impacts the user satisfaction~\cite{paterna2014ambient, egilmez2015user, park2018accurate}.
Therefore, mobile platforms must deliver the highest performance without violating 
thermal constraints to maximize user 
satisfaction~\cite{sahin2018maestro,prakash2016improving,
	muthukaruppan2013hierarchical}.

\vspace{1mm}
Competitive performance and fast response times are enabled 
by using multiple resources and operating them at high frequencies. 
For instance, the interactive governor on Android devices sets the frequency to the highest value whenever it detects user interactions. 
While this choice maximizes the response time and frame rate, 
it can also lead to thermal violations~\cite{brooks2007power,isci2006analysis,sahin2018maestro}. 
Higher chip temperature has a cascading effect of increasing the skin 
temperature, 
thus degrading the user experience~\cite{paterna2014ambient, egilmez2015user, park2018accurate}. 
Therefore, mobile platforms manage the temperature using thermal governors to mitigate this problem~\cite{prakash2016improving, sahin2018maestro, Henkel2019smart, 
hanumaiah2014steam}.

\vspace{1mm}
Thermal governors monitor the temperature at the critical hotspots on the chip. 
If the temperature of any hotspot exceeds the 
thermal limit, the governors react by decreasing the frequency of the processing elements. 
In extreme cases, the governors resort to powering the cores off to reduce the temperature of the device~\cite{bhat2018algorithmic,pagani2014tsp}. 
However, these actions can lead to significant performance 
degradation~\cite{sahin2016qscale,sahin2015impacts,wang2018optic,miyoshi2002critical}.
 
\vspace{1mm}
In this paper, we first present experimental case studies with Nexus 6P 
smartphone and most popular Android apps. 
This experimental study allows us to study the operation of the default thermal 
governors shipped with commercial phones. 
Our results show that thermal throttling degrades the performance by
as much as 34\% while running popular Android applications.  
Our empirical results also demonstrate the need for integrated thermal-frequency governors.
Thermal governors throttle the whole system instead of 
selectively throttling the resources that increase the temperature. 
For example, if a background application increases the temperature, 
the governors decrease the frequency of all processors in the system. 
This choice degrades the performance of all foreground applications.  
Finally, the outputs of the thermal and frequency governors may contradict each other~\cite{sahin2018maestro}. 
More specifically, frequency governors increase the operating frequency to ensure responsiveness, while thermal governors throttle frequencies to maintain the temperature within safe limits.  
Hence, we also propose an approach that can throttle select applications without 
affecting the performance of other apps in the system. 
We demonstrate our approach on the 
Odroid-XU3~\cite{ODROID_Platforms} board that employs a 
big.LITTLE heterogeneous system-on-a-chip, 
since it provides more flexibility to modify the default governors.


\vspace{1mm}
This paper is a part of the DATE 2019 Special Session on ``Smart Resource 
Management and Design Space Exploration for Heterogeneous Processors''. 
The other two papers of this special session are: ``Smart Thermal Management 
for Heterogeneous Multicores''~\cite{Henkel2019smart} and 
``Design and Optimization of Heterogeneous Manycore Systems enabled 
by Emerging Interconnect Technologies: Promises and 
Challenges''~\cite{Joardar2019hybrid}.
The rest of the paper is organized as follows. 
Section~\ref{sec:related_work} overviews the related work. 
In Section~\ref{case_study}, we perform a case study 
with the Nexus 6P phone to illustrate the behavior of default thermal governors. 
Section~\ref{odroid_study} presents a control algorithm 
that can selectively throttle applications 
and demonstrates it on the Odroid-XU3 board. 
Finally, Section~\ref{conclusions} 
concludes the paper with some future directions.

\section{Related Work} \label{sec:related_work}
A significant amount of recent research effort has focused on developing 
algorithms for dynamic frequency management under thermal 
constraints~\cite{bhat2018algorithmic,sahin2016qscale,
	prakash2016improving,sahin2018maestro,egilmez2015user}. 
For instance, the work in~\cite{sahin2016qscale,sahin2018maestro} proposes a 
closed-loop controller to maximize the quality of service~(QoS) provided by apps under thermal constraints. 
The controller ensures that apps can sustain their target QoS for a longer duration when compared to greedy approaches.
Similarly, the work in~\cite{prakash2016improving} considers a cooperative 
CPU-GPU governor for improving the performance of gaming apps.
Authors in~\cite{egilmez2015user} first develop a skin temperature predictor 
for smartphones
and then use it for user-specific skin temperature aware frequency and voltage 
selection.
The algorithm proposed in~\cite{bhat2018algorithmic} uses a temperature 
prediction to determine a power budget that does not violate the thermal 
constraint. Then, it uses a gradient search algorithm to reduce the frequency 
of resources that result in the minimum performance loss.
While these algorithms consider dynamic frequency management under thermal 
constraints, they do not consider the problem of selectively throttling 
background apps without affecting the foreground apps.
%

\section{Empirical Study of Thermal Throttling on Commercial Smartphone} \label{case_study}
\subsection{Experimental Setup}
We perform our empirical study on a Nexus 6P~\cite{nexus6P} smartphone
that runs Android 7 OS. 
A commercial smartphone with actual form-factor is preferred over commonly used 
experimental boards since the form-factor and packaging affect the thermal 
behavior. 
Furthermore, it enables us to evaluate the temperature and power management governors shipped to the end users. 
The Nexus 6P smartphone houses a Qualcomm Snapdragon 810 system-on-a-chip (SoC)~\cite{Qual_Snap810}, 
which integrates four low-power Cortex-A53 cores, 
four high-performance Cortex-A57,  
and an Adreno 430 GPU.
The phone also has sensors to measure the temperature at various points on the 
SoC, such as the chip package, memory, and flash memory. 
Among these, we measure the temperature at the chip package, since it is used 
by the default governors to make thermal management decisions.
In contrast to evaluation boards that include power measurement sensors, 
the Nexus 6P phone does not include power measurement sensors. 
Therefore, we use a National Instruments data acquisition system~\cite{PXIe-4081} to measure the power 
consumption of the phone
at a sampling rate of 1~KHz. 

\vspace{1mm}
\noindent \textbf{Applications under study:} 
Using our setup, we evaluate the power, performance, 
and thermal behavior of the most popular apps on the Google play store.
We present the results for five representative apps 
from the top 30 apps on the Google play store.
More specifically, we choose two games, one shopping app, one 
video conferencing app and one social media app, since they are the most 
popular app categories~\cite{Statista2018_app_categories}. 

We run each of these apps under two scenarios. 
First, we disable the default temperature governor to measure the baseline performance of each app. 
Then, we repeat the experiments by enabling the default temperature governor shipped with the phone. 
This controlled experiment allows us to measure the impact on performance due to thermal throttling. 
In the following section, we analyze the temperature and performance of each application.

\subsection{Performance and temperature analysis of popular apps}

\begin{figure}[b]
	\centering
		\vspace{-5mm}
	\includegraphics[width=0.95\linewidth]{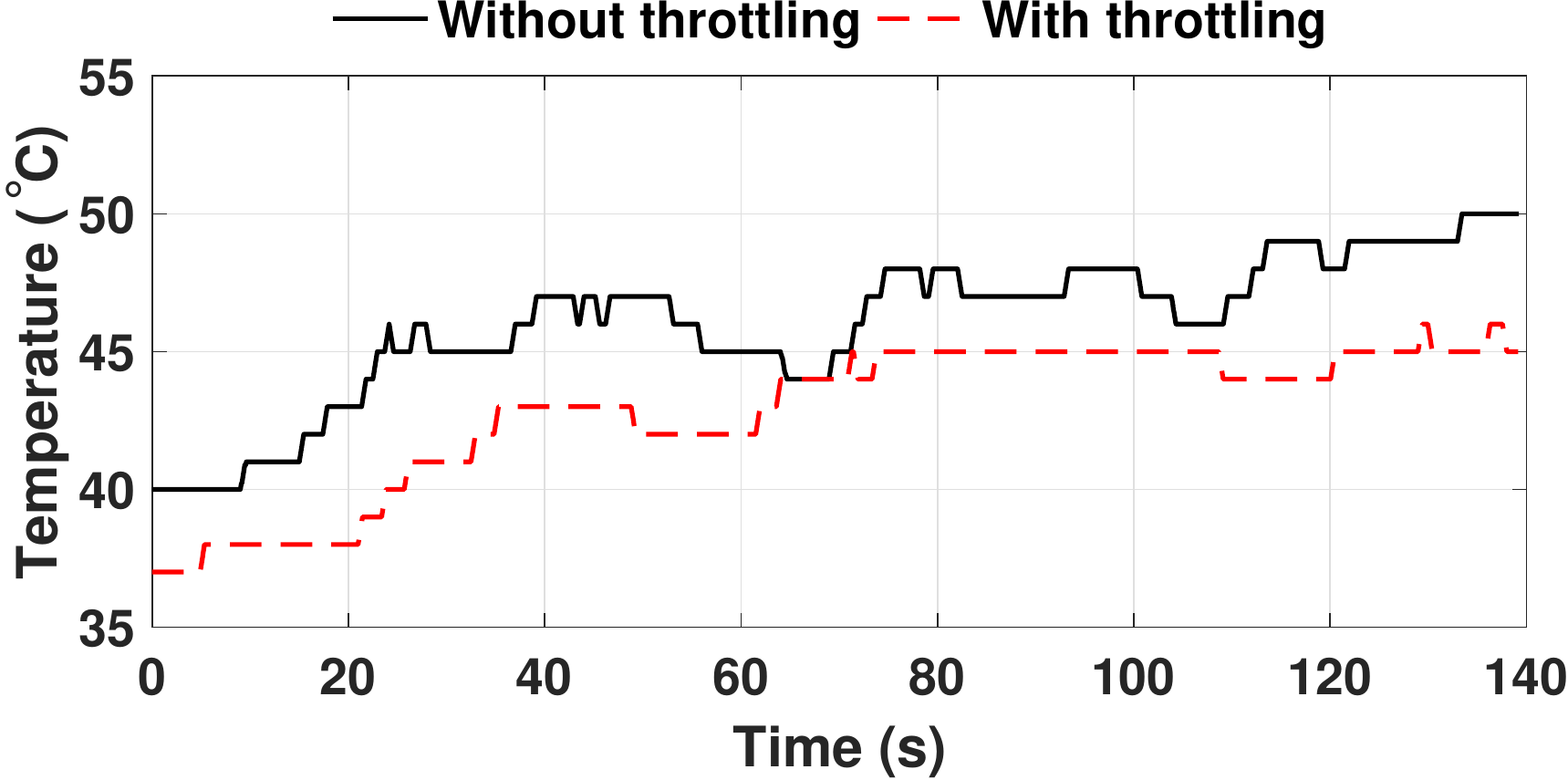}
		\vspace{-2mm}
	\caption{Temperature profile for Paper.io game.}
	\label{fig:paper_io}
	\includegraphics[width=1\linewidth]{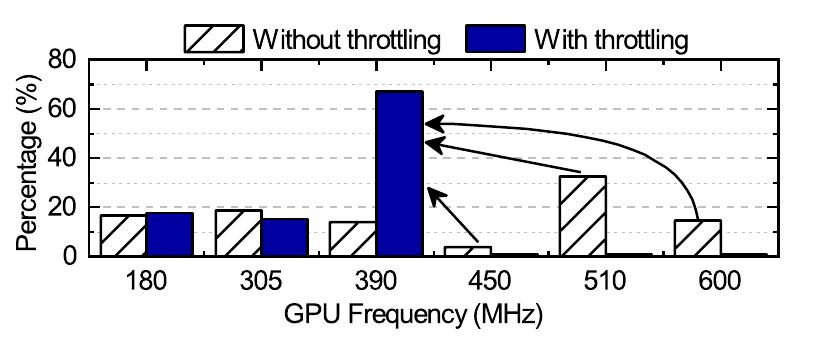}
	\vspace{-7mm}
	\caption{Usage of GPU frequencies in the Paper.io game}
	\label{fig:paper_io_throttling}
\end{figure}

\noindent\textbf{Paper.io game:} 
Paper.io is one of the top five games available in the Google play store.
Figure~\ref{fig:paper_io} shows the behavior 
of the temperature of the phone when running the Paper.io game. 
We observe that there is a significant increase in 
the temperature when we disable thermal throttling. 
The package temperature reaches 50$\degree$C at the end of the run. 
Then, we enable the default thermal governor and re-run the game. 
The dashed red line in Figure~\ref{fig:paper_io} shows that the thermal governor can successfully control the temperature.  
However, it comes with a significant performance penalty of about 12 frames per second (\textit{FPS}). 
To understand the behavior of the phone in more detail, 
we analyze the usage of different GPU frequencies in Figure~\ref{fig:paper_io_throttling}. 
When there is no throttling,  
the GPU operates 32\% of the time at 510~MHz and 15\% of the time at 600~MHz, which are the two highest frequencies. 
Figure~\ref{fig:paper_io_throttling} clearly shows that throttling decreases the use of these frequencies significantly. 
More specifically, the use of 510~MHz and 600~MHz drop to zero, 
while the percentage of operation at 390~MHz increases sharply 
from 15\% to 67\%. 
Since the GPU spends noticeably more time at lower frequencies, 
the game suffers a drop in performance. 
More specifically, the frame rate drops from 35 \textit{FPS} to 23 \textit{FPS}, 
which corresponds to 34\% degradation in performance, 
as summarized in Table~\ref{tab:fps_comparison}.

\begin{table}[t]
	\centering
	\vspace{1mm}
	\caption{\small Median frame rate achieved while running popular Android apps with and without throttling, respectively.}
	\label{tab:fps_comparison}
	\begin{tabular}{@{}lccc@{}}
		\toprule
		& \multicolumn{2}{c}{Frame Rate} & \\
		App & \begin{tabular}[c]{@{}c@{}}Without \\ Throttling\end{tabular} 
		& \begin{tabular}[c]{@{}c@{}}With \\ Throttling\end{tabular} & 
		\begin{tabular}[c]{@{}c@{}} Percentage Reduction \end{tabular} \\ 
		\midrule
		Paper.io & 35 \textit{\textit{FPS}} & 23 \textit{\textit{FPS}} & 34\% \\
		Stickman Hook & 59 \textit{FPS}& 40 \textit{FPS}& 32\% \\
		Amazon & 35 \textit{FPS}& 28 \textit{FPS}& 20\% \\ 
		Google Hangouts & 42 \textit{FPS}& 38 \textit{FPS}& 10\% \\ 
		Facebook & 35 \textit{FPS}& 24 \textit{FPS}& 31\% \\ \bottomrule
	\end{tabular}
	\vspace{-5mm}
\end{table}

\begin{figure}[b]
	\centering
	\vspace{-4mm}
	\includegraphics[width=0.95\linewidth]{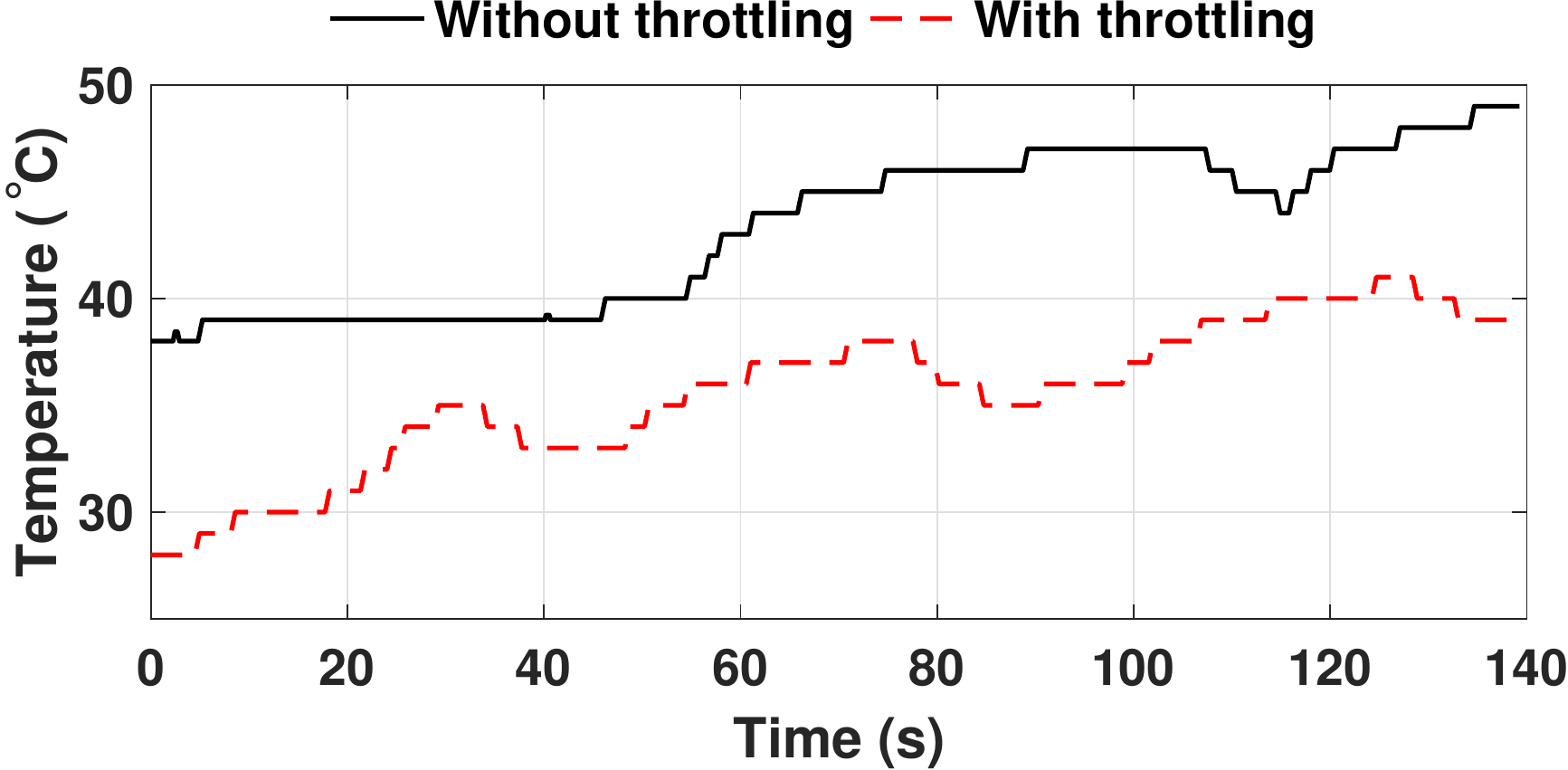}
	\caption{Temperature profile for Stickman Hook game.}
	\label{fig:stickman}
	\includegraphics[width=1\linewidth]{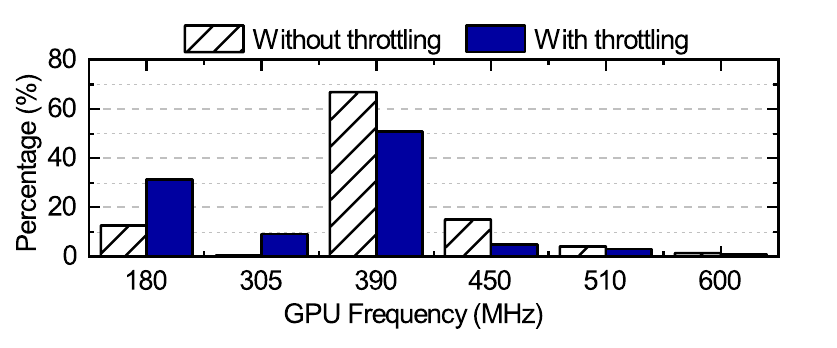}
	\caption{Usage of GPU frequencies in the Stickman Hook game.}
	\vspace{-2mm}
	\label{fig:stickman_throttling}
\end{figure}

\vspace{1mm}
\noindent\textbf{Stickman Hook game:}
Stickman Hook is another game that is in the top 
five apps available on the Google play store.
Figure~\ref{fig:stickman_throttling} shows the temperature 
profile of the phone while playing Stickman Hook with the default thermal 
governor and by disabling it. 
We observe that the temperature reaches significantly higher values, 
especially after running the application for 50 seconds, when thermal throttling is disabled. 
As expected, thermal throttling enables lower temperatures 
and keeps the maximum temperature below 40$\degree$C. 
Again, this comes at the expense of the performance. 
Figure~\ref{fig:stickman_throttling} shows that 
the use of 450~MHz and 510~MHz drops almost to zero. 
Furthermore, the operation at 390~MHz drops from 67\% to 51\%. 
In contrast, the operation at the lowest (180~MHz) and the
second lowest frequency (305~MHz) 
increase from 12\% to 31\% and 0\% to 9\%, respectively.  
Consequently, operating at lower frequencies leads to a 32\% drop in the median \textit{FPS}, as shown in Table~\ref{tab:fps_comparison}.

\begin{figure}[t]
	\centering
	\includegraphics[width=0.95\linewidth]{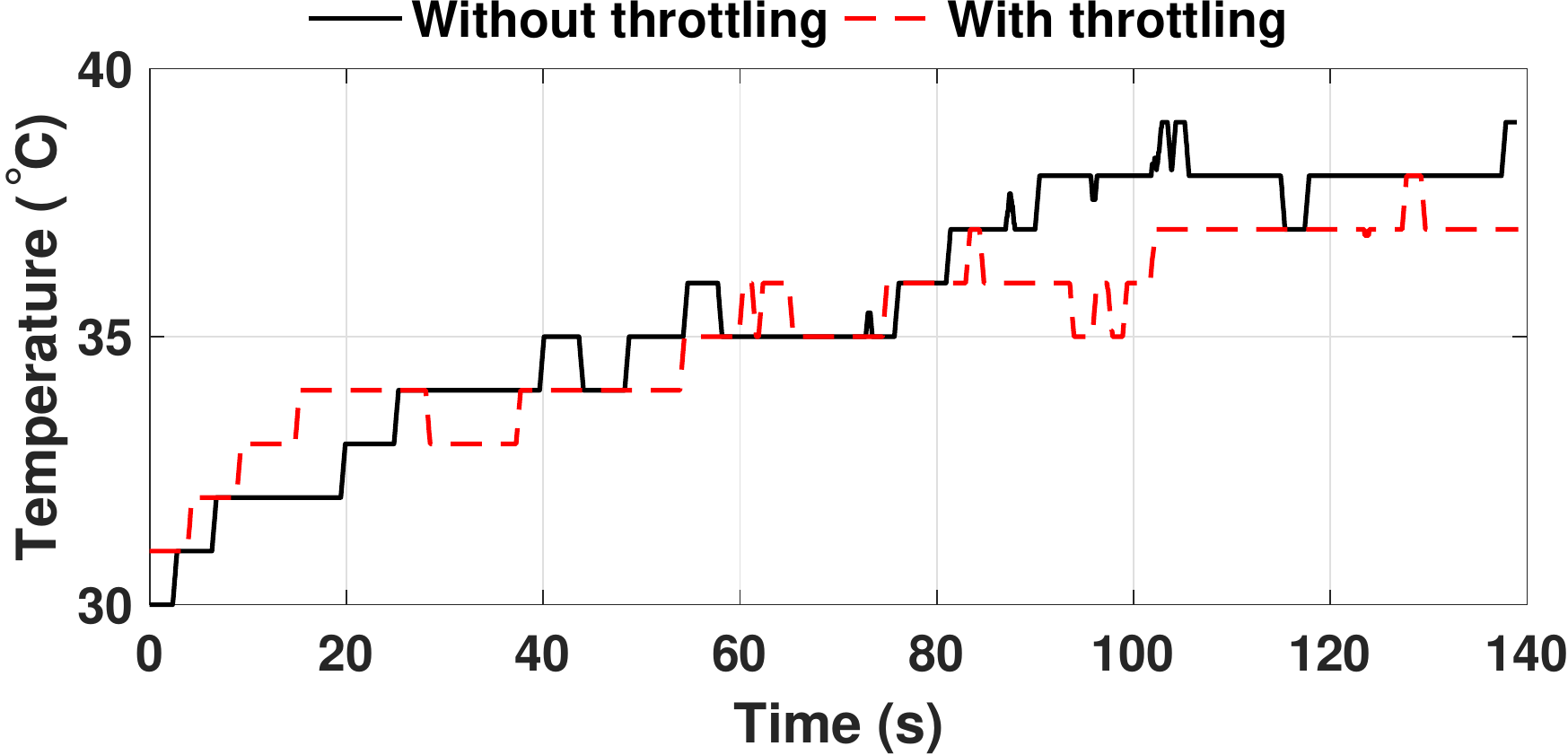}
	\vspace{-1mm}
	\caption{Temperature profile for Amazon 
		shopping app.}
	\label{fig:amazon}
	\vspace{1mm}
	\includegraphics[width=1\linewidth]{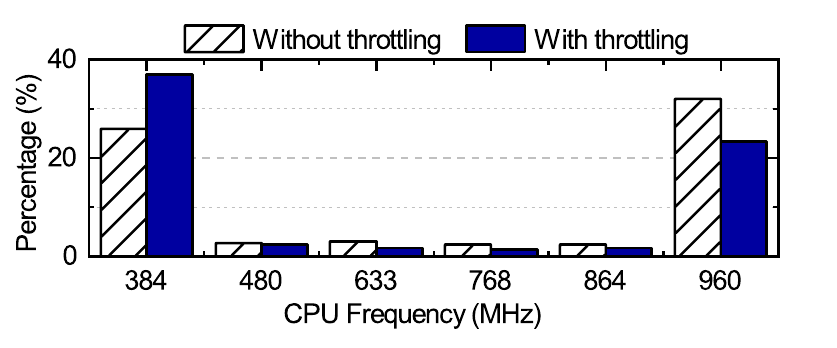}
	\vspace{-6mm}
	\caption{Usage of big core frequencies in the Amazon app.}
	\label{fig:amazon_throttling}
	\vspace{-5mm}
\end{figure}

\begin{figure*}[b]
	\centering
	\vspace{-2mm}
	\includegraphics[width=1\linewidth]{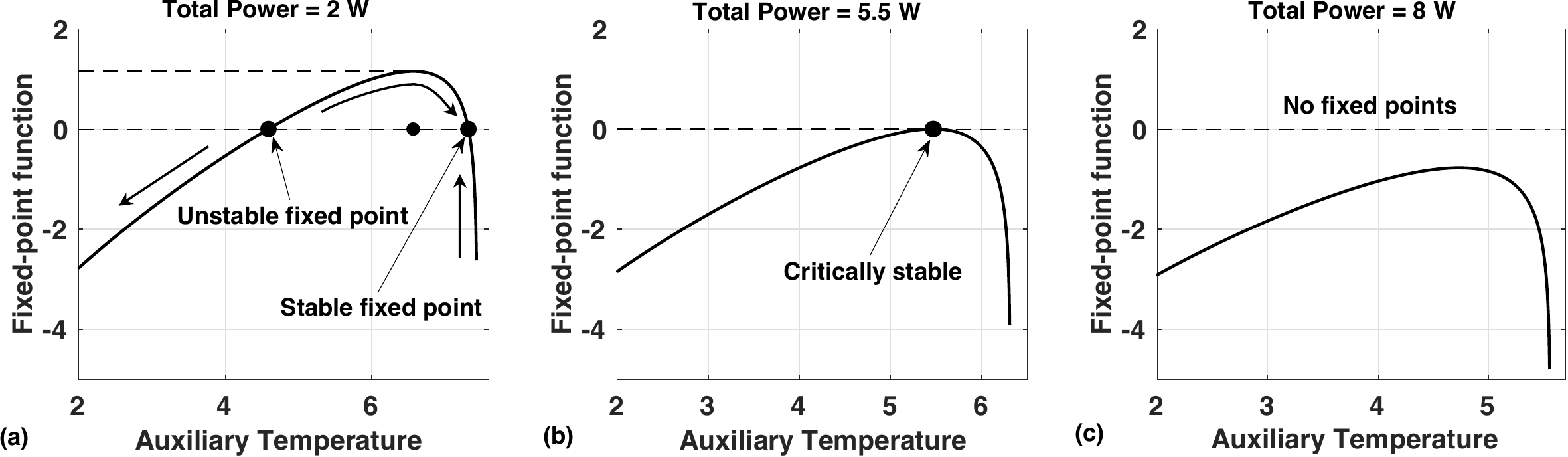}
	\caption{Illustration of the fixed point functions for three power 
		consumption values.}
	\label{fig:fixed_point_illustrate}
\end{figure*}

\noindent\textbf{Amazon shopping:}
Amazon is a popular online shopping app available on the Google play store. 
In contrast to the gaming apps, it primarily uses the CPU when it is active.
Figure~\ref{fig:amazon} shows the behavior of the temperature of the phone when 
using the Amazon app. 
We observe that the temperature with and without 
throttling closely follow each other during the first 80~seconds of use. 
After that, the temperature increases if the thermal throttling is disabled. 
In contrast, the thermal governor is able to maintain the temperature by reducing the CPU frequency. 
The reduction in CPU frequency can be analyzed using the data in Figure~\ref{fig:amazon_throttling}. 
When there is no throttling, the CPU operates 32\% of the time at 960~MHz. 
This percentage drops to 23\% with throttling. 
In contrast, the operating at the lowest frequency (384~MHz) increases from 
25\% to 37\% with throttling. 
As a result, the frame processing rate drops from 35~\textit{FPS} to 
28~\textit{FPS} similar to the other applications, as shown in 
Table~\ref{tab:fps_comparison}.

\vspace{1mm}
\noindent \textbf{Google Hangouts and Facebook apps:} 
In addition to the previous apps, we also analyze the performance of Google Hangouts 
and Facebook apps.
We do not plot the temperature profiles for these two apps since they are similar to that of the Amazon app. 

Consistent with the other apps, we observe that the default governor reduces the temperature at the expense of the frame processing rate. 
In particular, thermal throttling reduces the frame rate from 42~\textit{FPS} 
to 38~\textit{FPS} while running Google Hangouts. 
Similarly, the frame of Facebook drops from 35~\textit{FPS} to 24~\textit{FPS} 
while playing a game in the app.

The experiments with popular Android apps show that 
thermal governors regulate the temperature 
at the cost of performance 
since they \textit{react} to temperature violations by throttling the frequency 
of all the resources in the system. 
Theoretical analysis of power-temperature dynamics 
can help in predicting potential violations before they 
happen~\cite{bhat2018algorithmic,cochran2013thermal,prakash2016improving}.
Furthermore, they can guide the utilization of different resources, 
such as big versus little cores, judiciously to prevent temperature violations
with minimal impact on performance.
The next section illustrates this idea with a graphical analysis 
and empirical results.

\section{Thermal Management using Power-Temperature Stability Analysis} 
\label{odroid_study}

\subsection{Analysis of the Power-Temperature Dynamics}
\label{sec:power-thermal-analysis}
Power and temperature form a well-known positive feedback 
system~\cite{bhat2017power,liao2003coupled}. 
The junction temperature increases with power consumption. 
Higher temperature, in turn, leads to a 
higher leakage power, which contributes to a further increase in the 
temperature~\cite{brooks2007power}.
The temperature converges to a stable fixed point when the system is stable.  
In contrast, an unstable system experiences a thermal runaway. 
The stability of power-temperature dynamics depends on the power consumption that changes at runtime.  
Therefore, it is imperative to analyze the stability of the dynamics at runtime.


A theoretical analysis of the power-temperature dynamics is presented in~\cite{bhat2017power}. 
This analysis enables us to derive the sufficient and necessary conditions for the existence of temperature fixed points. 
The fixed points are the \textit{roots} of a concave function of an auxiliary temperature, as illustrated in Figure~\ref{fig:fixed_point_illustrate}. 
The auxiliary temperature is inversely proportional to the actual temperature in Kelvin. 
Therefore, a higher auxiliary temperature corresponds to a lower temperature and vice versa. 
This section presents a graphical analysis of the stability, while the complete 
theoretical proof is presented in~\cite{bhat2017power}. 

Figure~\ref{fig:fixed_point_illustrate}a plots the fixed-point function when 
the power consumption is 2~W using parameters obtained for Odroid XU3.  
We observe that the fixed-point function is indeed concave. 
Furthermore, it has two roots that correspond to two temperature fixed points.
The arrows along the plot show that the auxiliary temperature 
\textit{decreases} with each fixed-point iteration when the function is 
\textit{negative}. 
In contrast, it \textit{increases} at each iteration when the function is \textit{positive} (i.e., between two roots).  
Hence, the larger root attracts the temperature trajectories to itself, 
i.e., it is the stable fixed point, as illustrated in 
Figure~\ref{fig:fixed_point_illustrate}a. 
In contrast, the temperature diverges from the unstable fixed point. 
Hence, any trajectory that starts between the roots will converge to the stable fixed point. 
If the initial point is to the left of the unstable fixed point, there is a thermal runaway.

When the power consumption increases, 
the fixed-point function moves down, as shown in 
Figure~\ref{fig:fixed_point_illustrate}b. 
The power-temperature dynamics continue to have two fixed-points until the 
power consumption reaches a critical value. 
In our example, the roots of the fixed-point function converge, 
i.e., there is only one root, when the power consumption reaches 5.5~W. 
Any further increase in the power consumption results in an unstable system with no 
fixed points. For instance, Figure~\ref{fig:fixed_point_illustrate}c plots 
the fixed-point function when the total power is 8~W. We see that it does not 
intersect the x-axis, which shows that the system does not have any fixed points.
In summary, we can determine the stability of the power-temperature 
dynamics by looking at the number of roots of the fixed-point function.
In the next section, we use the thermal stability analysis to design an 
application-aware thermal management algorithm.

\subsection{Application-Aware Thermal Management}

The thermal stability analysis overviewed in the previous section 
provides an efficient method to calculate the steady-state temperature~(fixed-point temperature) of the system as a function of the power consumption. 
We also use it to estimate the time that will pass to reach the fixed point. 
This information can be used by a governor to enable 
a new class of  dynamic thermal and power management~(DTPM) algorithms. 
As illustrated in Section~\ref{case_study}, thermal governors in current mobile platforms typically throttle 
the frequency of all active cores when a thermal violation is detected. 
Instead of degrading the performance of all active applications in 
the system, we utilize the fixed-point predictions to design an application-aware 
thermal management algorithm.

We start by using the thermal stability analysis to 
determine the stable fixed-point temperature. 
If this temperature exceeds a specified thermal limit, there may be a thermal violation in the future.  
Therefore, the algorithm estimates the time it will take for the system to 
reach the fixed point. If the algorithm detects that the time to reach the 
fixed-point is less than a user-defined limit, it means that there is an 
imminent possibility of a thermal violation. 
In this case, the algorithm finds the process that has the highest power 
consumption by monitoring the average utilization of each active process for a one-second window. 
We use a window to filter out momentary peaks in the 
power consumption.
Finally, the algorithm moves the most power-hungry process to low power processors. 
We repeat this process every 100 ms to capture runtime variations in the system activity effectively. 

The primary advantage of the proposed algorithm is penalizing only the process that causes a thermal violation. 
The other processes continue operating at maximum performance in strong contrast to the default governors, which throttle the whole system to regulate the temperature. 
The algorithm also lets processes with real-time requirements register themselves so that they are not penalized.
We present an empirical evaluation of the proposed algorithm in the next section.




\subsection{Experimental results on Odroid-XU3}

This section evaluates the proposed application-aware thermal management algorithm on the Odroid XU3 board. 
The board employs the Samsung Exynos 5422 
SoC~\cite{ODROID_Platforms}, which integrates four Cortex-A15~(big) cores, four Cortex-A7 cores, and a Mali T628 
GPU. In addition to the processing elements, the board includes thermal sensors 
to measure the temperature of each big core and the GPU. It also provides 
current sensors to measure the power consumption of the little cluster, big 
cluster, main memory, and the GPU. 
We run Android 7.1 with Linux kernel 3.10.9. 
The governor is invoked every 100 ms, 
along with the default frequency governors. 
We evaluate the proposed controller with commonly used 3DMark and Nenamark benchmarks. 
\textcolor{black}{We use the Odroid XU3 board to illustrate the proposed algorithm since it provides a higher level of control to modify the frequency and thermal governors in the system. Furthermore, the Odroid board provides individual power sensors that allow for a better evaluation of the proposed algorithm.}

To evaluate the algorithm, we run a real-time GPU 
benchmark, along with a computationally intensive task in the background. 
The default policy is to use the thermal management policy in the Linux 
kernel~(3.10.9). Specifically, it uses thermal trip points and ARM 
intelligent power allocation algorithm to control the 
temperature~\cite{armIPA}. 
During these experiments, we disable the fan on the board 
since it is not feasible for mobile platforms.

\begin{figure}[b]
	\centering
	\includegraphics[width=1\linewidth]{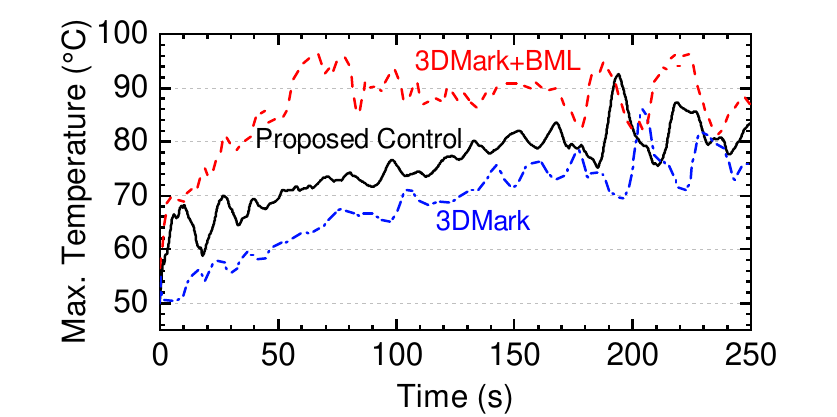}
	\vspace{-3mm}
	\caption{The maximum temperature of the system when running 3DMark application}
	
	\label{fig:csm_control}
\end{figure}
\begin{figure}[t]
	\centering
	\includegraphics[width=1\linewidth]{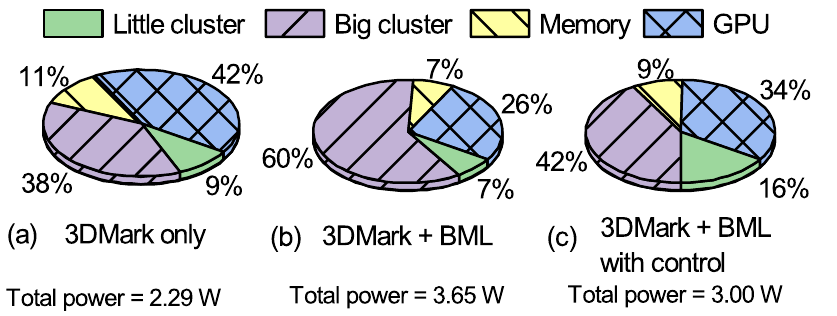}
	\caption{Power consumption distribution of 3DMark }
	\label{fig:3dmark_power_distrubition}
	\vspace{-0mm}
\end{figure}

\vspace{1mm}
\noindent \textbf{3D Mark alone with the default governor:} 
First, we run the 3DMark benchmark alone without any background application. 
This experiment gives an upper bound for the performance.
It also gives the baseline temperature profile, which is shown by the blue curve in Figure~\ref{fig:csm_control}. 
The corresponding power consumption and its distribution among the major components on the SoC are shown in Figure~\ref{fig:3dmark_power_distrubition}a.  
We observe that the GPU has the highest power 
consumption, followed by the big core cluster. 
This is expected since 3DMark is a GPU-heavy application. 

\vspace{1mm}
\noindent \textbf{3D Mark + BML with the default governor:} 
Next, we re-run the 3DMark benchmark while also executing the basicmath large~(BML) application~\cite{guthaus2001mibench} in the background. 
As expected, the BML application increases the power consumption of the big core, and consequently the whole board. 
More specifically, the total power consumption jumps to 3.65~W, 
and the contribution of the big core cluster increases from 38\% to 60\%, 
as depicted in Figure~\ref{fig:3dmark_power_distrubition}b. 
Consequently, larger power consumption leads to a higher temperature, 
as shown by the red dashed line in Figure~\ref{fig:csm_control}. 
As a result, the thermal governor starts to throttle the frequency of the system. Throttling of all the resources 
causes a drop in the performance of 3DMark Graphics Test 1~(GT1) from 97 to 86 
\textit{FPS}, as shown in the third column of 
Table~\ref{table:perf_comparison}. 
Similarly, the performance of Graphics Test 2~(GT2) also decreases 
from 51~\textit{FPS} to 49~\textit{FPS}.

\begin{table}[b]
	\centering
	\caption{Comparison of application performance with the proposed control 
		algorithm}
	\label{table:perf_comparison}
	\begin{tabular}{@{}lccc@{}}
		\toprule
		Test                & App. Alone & App. + BML & 
		\begin{tabular}[c]{@{}c@{}}App. + BML with\\ Proposed 
			Control\end{tabular} \\ \midrule
		3DMark GT1          & 97 \textit{FPS}   & 86 \textit{FPS}   & 93
		\textit{FPS}                                                            
		\\
		\midrule
		3DMark GT2          & 51 \textit{FPS}   & 49 \textit{FPS}   & 51 
		\textit{FPS}                                                            
		
		\\
		\midrule
		Nenamark3           & 3.5 levels & 3.4 levels & 3.5 
		levels                                                                 
		\\
		\bottomrule
	\end{tabular}
\end{table}

\vspace{2mm}
\noindent \textbf{3D Mark + BML with the proposed controller:} 
Finally, we apply the proposed control algorithm when running 3D Mark and BML together. 
As in the previous experiment, the BML application causes the 
temperature to rise. 
However, the algorithm detects and migrates the background application to the little 
cluster as soon as it predicts a thermal violation. 
This enables the proposed algorithm to control the temperature of the system, 
as shown using a black line in~Figure~\ref{fig:csm_control}. The migration also
causes a reduction in the big core power consumption from 60\% to 42\%, as 
seen by the pie chart in Figure~\ref{fig:3dmark_power_distrubition}c. 
We see a corresponding increase in the power consumption of the little core from 7\% to 16\% since the BML application is running on the little cores.
The migration is able to effectively throttle the BML 
application without affecting the 3DMark application. 
This can be observed in the last column of Table~\ref{table:perf_comparison}, 
where the performance loss in 3DMark GT1 and GT2 are minimal.

The same steps are repeated with the Nenamark benchmark in order to test the 
controller on a workload with different characteristics.
Nenamark measures the number of levels of the benchmark that can be run at a 
given frame rate. The benchmark terminates when the frame rate drops below the 
desired level.
The performance achived while running Nenamark under the three scenarios is summarized in the last row of Table~\ref{table:perf_comparison}. 
We observe that the number of levels which can run at the desired frame rate is lower when the whole system is throttled. 
However, with the proposed control the number of levels is equal to 
the baseline performance of the application without any background applications.
In summary, the proposed control algorithm is able to effectively migrate the 
power-hungry app without causing any adverse effect on the main app.


\vspace{-1mm}
\section{Conclusions}\label{conclusions}
This paper presented experiments to study the power, performance and thermal 
behavior of modern smartphones. Specifically, we performed experiments on the 
Nexus 6P phone with 
popular apps to understand the loss in their performance due to thermal 
throttling. Using insights from these experiments and our previous work, we 
developed an algorithm that can selectively throttle background apps without 
affecting foreground apps. 
Experiments on the Odroid-XU3 board show that the algorithm can throttle 
background apps effectively.
The experimental case study in this paper can be used as a baseline when 
evaluating future thermal management algorithms. Furthermore, it can be used by 
application developers to optimize their apps such that they do not experience 
thermal throttling.

\vspace{2mm}
\noindent\textbf{Acknowledgments:}
This work was supported in part by National Science Foundation (NSF) grant 
CNS-1526562 and Semiconductor Research Corporation (SRC) task 
2721.001.

\vspace{-1mm}
\footnotesize{\bibliographystyle{IEEEtranS}}



\end{document}